\documentclass[preprint,review, 12pt]{elsarticle}
 
\usepackage{amssymb}

\usepackage{subfigure}
\usepackage{textcomp}
\usepackage{lineno}
\usepackage{graphicx}
\usepackage{subfigure}
\usepackage{multirow}
\usepackage{amsmath}
\usepackage{amssymb}
\usepackage{ulem}
\usepackage{verbatim}
\usepackage{upgreek}
\usepackage{xcolor}
\usepackage{makecell}
\usepackage{threeparttable}
\usepackage[colorlinks,
            linkcolor=blue,
            anchorcolor=blue,
            citecolor=blue,
            urlcolor=blue]{hyperref}
\usepackage{tablefootnote}

\graphicspath{{fig/}}
\newif\ifcomment
\commenttrue
\commentfalse

\ifcomment
\usepackage[UTF8]{ctex}
\linenumbers
\fi

\usepackage{xspace} 
\makeatletter
\ifcase \@ptsize \relax
  \newcommand{\miniscule}{\@setfontsize\miniscule{4}{5}}
\or
  \newcommand{\miniscule}{\@setfontsize\miniscule{5}{6}}
\or
  \newcommand{\miniscule}{\@setfontsize\miniscule{5}{6}}
\fi
\makeatother
\DeclareRobustCommand{\optbar}[1]{\shortstack{{\miniscule (\rule[.5ex]{0.75em}{.18mm})}
  \\ [-.7ex] $#1$}}
\def\porpbar    {\kern 0.18em \optbar{\kern -0.18em p}{}\xspace}

\journal{NIMA}

\begin{document}
\begin{frontmatter}

\title{\boldmath Performance studies of a SiPM-readout system with a pico-second timing chip}

\author[a,b]{Xin~Xia}
\author[a,b]{Dejing~Du}
\author[a,b]{Xiaoshan~Jiang}
\author[a,b]{Yong~Liu\corref{cor1}}
\ead{liuyong@ihep.ac.cn}
\author[a]{Bo~Lu\corref{cor1}}
\ead{luboihep@outlook.com}
\author[a,b]{Junguang~Lyu}
\author[a,b]{Baohua~Qi}
\author[a,b]{Manqi~Ruan}
\author[a,b]{Xiongbo~Yan}
\cortext[cor1]{Corresponding authors.}

\address[a]{Institute of High Energy Physics, Chinese Academy of Sciences, 19B Yuquan Road, Shijingshan District, Beijing 100049, China}
\address[b]{University of Chinese Academy of Sciences, 19A Yuquan Road, Shijingshan District, Beijing 100049, China}

\begin{abstract}
A pico-second timing (PIST) front-end electronic chip has been developed using $55~\mathrm{nm}$ CMOS technology for future electron-positron collider experiments (namely Higgs factories). Extensive tests have been performed to evaluate the timing performance of a dedicated SiPM-readout system equipped with a PIST chip. The results show that the system timing resolution can achieve $45~\mathrm{ps}$ for SiPM signals at the minimum-ionizing particles (MIP) level ($200~\mathrm{p.e.}$) and better than $ 10~\mathrm{ps}$ for signals larger than $1200~\mathrm{p.e.}$, while the PIST intrinsic timing resolution is $4.76 \pm 0.60~\mathrm{ps}$. The PIST dynamic range has been further extended using the time-over-threshold (ToT) technique, which can cover the SiPM response spanning from $\mathrm{\sim 900~p.e.}$ to $~\mathrm{\sim 40000~p.e.}$.

\end{abstract}


\end{frontmatter}


\section{Introduction}
\label{sec:intro}

To precisely measure the properties of the Higgs, W, Z bosons and explore new physics beyond the Standard Model, fast timing performance is crucial for the calorimetry of future electron-positron colliders, such as CEPC~\cite{CEPCStudyGroup:2018ghi}, FCC-ee~\cite{FCC:2018evy}, ILC~\cite{Behnke:2013lya}, and CLIC~\cite{Linssen:2012hp}. High-precision time-of-flight (ToF) measurements provided by electromagnetic calorimetry can complement $dE/dx$ measurements and significantly improve the particle identification (PID) performance with a required TOF resolution of around $50~\mathrm{ps}$~\cite{An:2018jtk}. In addition, timing measurements with a resolution at nanosecond-level can also enhance the hadronic energy resolution by approximately 3\% to 4\% through the local software compensation for the CALICE Analog Hadron Calorimeter (AHCAL)~\cite{Graf:2022lwa}.

\begin{figure*}
\centering
\includegraphics[height=.27\textwidth]{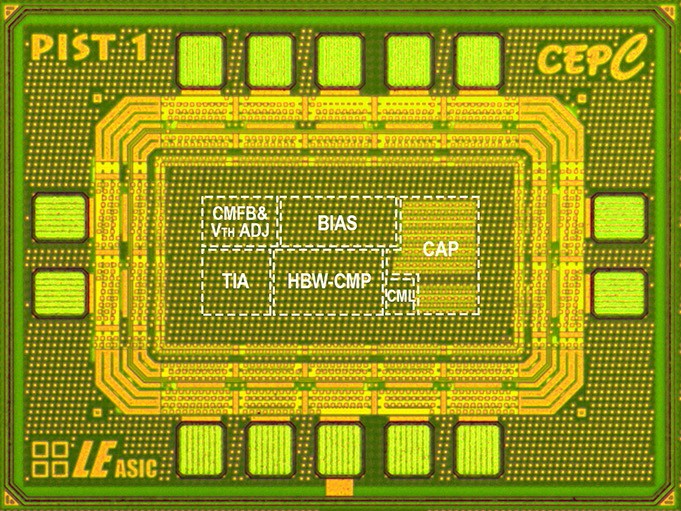}\quad
\includegraphics[height=.27\textwidth]{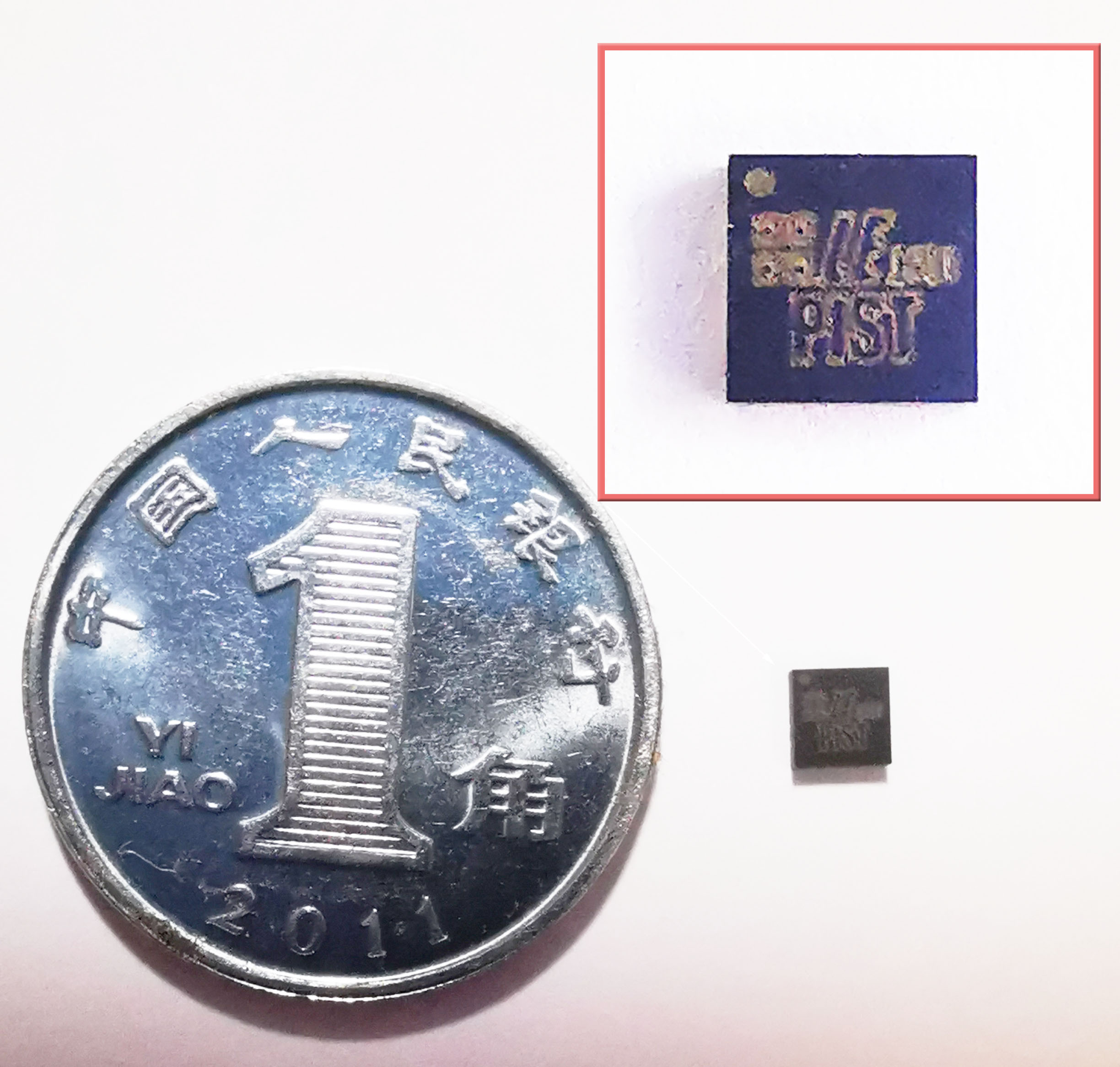}\quad
\includegraphics[height=.27\textwidth]{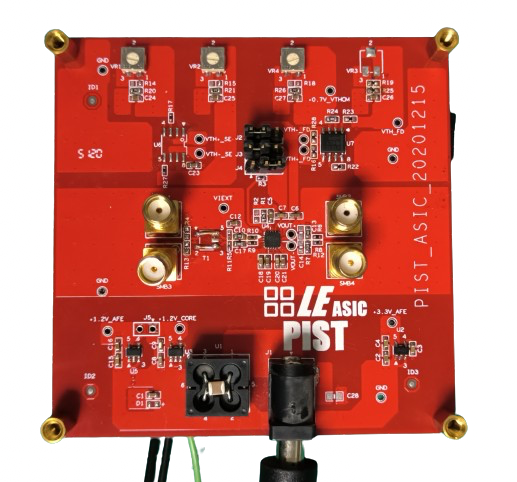}
\caption{The PIST die micrograph (left), a packaged PIST chip (middle) and an evaluation board (right) equipped with a PIST chip.}
\label{fig:PIST_photo}
\end{figure*}

Considering the importance of high precision time information, a pico-second timing (PIST) application-specific integrated circuit (ASIC) has been developed for processing SiPM signals. It utilizes $55~\mathrm{nm}$ CMOS technology, operates with a single $1.2~\mathrm{V}$ supply voltage, and has a low power dissipation of $15~\mathrm{mW}$ for a single channel~\cite{Lu:2023chl}. The photos of a die micrograph, a packaged PIST chip, and an evaluation board are shown in Fig.~\ref{fig:PIST_photo}. The chip is packaged in a 16-pin QFN (quad flat no-lead package) in an area of $3\times3~\mathrm{mm^{2}}$ with an additional exposed pad. Its core, including all the critical blocks, occupies an area of $220\times120~\mathrm{\mu m^{2}}$. Based on the $55~\mathrm{nm}$ CMOS technology for better immunity to total ionizing dose effects and a lower power consumption, the PIST chip can provide a wide-band amplification to accommodate the steep rising edge of SiPM signals for an excellent time resolution. It is operated in a low-noise mode to further improve the time resolution and also capable to perform energy measurements in a certain range through the ToT technique, with adjustable thresholds to enhance linearity.

The previous work by Lu et al.~\cite{Lu:2023chl} has elucidated the ASIC's principles and reported results with a clean single-ended square-wave voltage signal generated by a signal generator. In our study, we further investigated the combined PIST-SiPM timing performance by developing dedicated test stands and comprehensive characterisations of the ASIC's response to real SiPM signals.

This paper is organized as follows. Sec.~\ref{sec:system} introduces the experimental setups and analysis methodology for quantifying the timing performance and ToT response of the test stands, followed by Sec.~\ref{sec:results} and Sec.~\ref{sec:discussion} with measurement results and discussions, respectively. Conclusions and prospects are covered in Sec.~\ref{sec:conclusion}.

\section{Experimental setups and an analysis methodology}
\label{sec:system}
We have developed two setups and an analysis methodology accordingly to characterize the timing performance and ToT response of the PIST ASIC to the SiPM, with Sec.~\ref{sec:setups} introducing the design of the setups and characterization of the components, and Sec.~\ref{Sec:Ana} presenting the algorithm for quantifying the performance.
\subsection{Setups design and components characterization}
\label{sec:setups}
\begin{figure}
\centering
    \includegraphics[height=5cm]{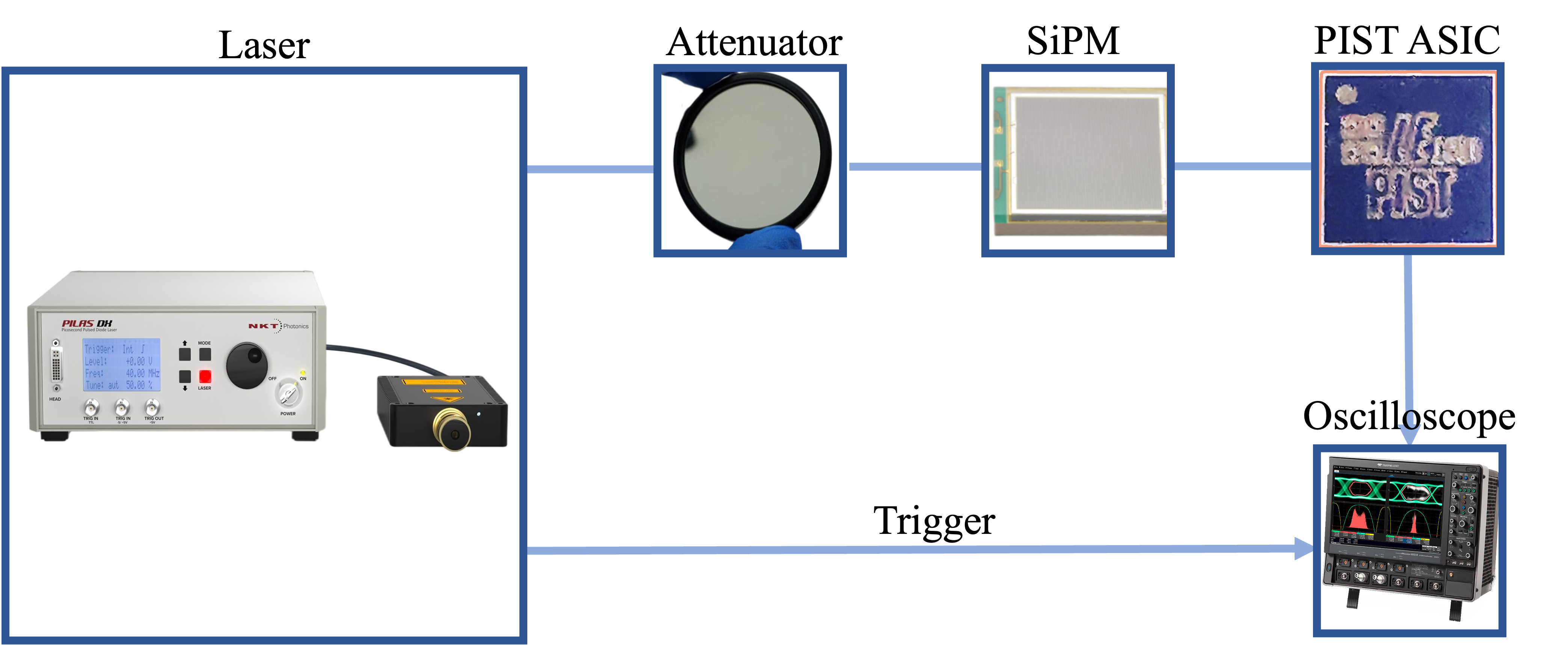}
\caption{The setup of the ASIC testing system (SiPM+PIST System).}
\label{fig:SiPM_PIST_System}
\end{figure}

\begin{figure}
\centering
    \includegraphics[height=5cm]{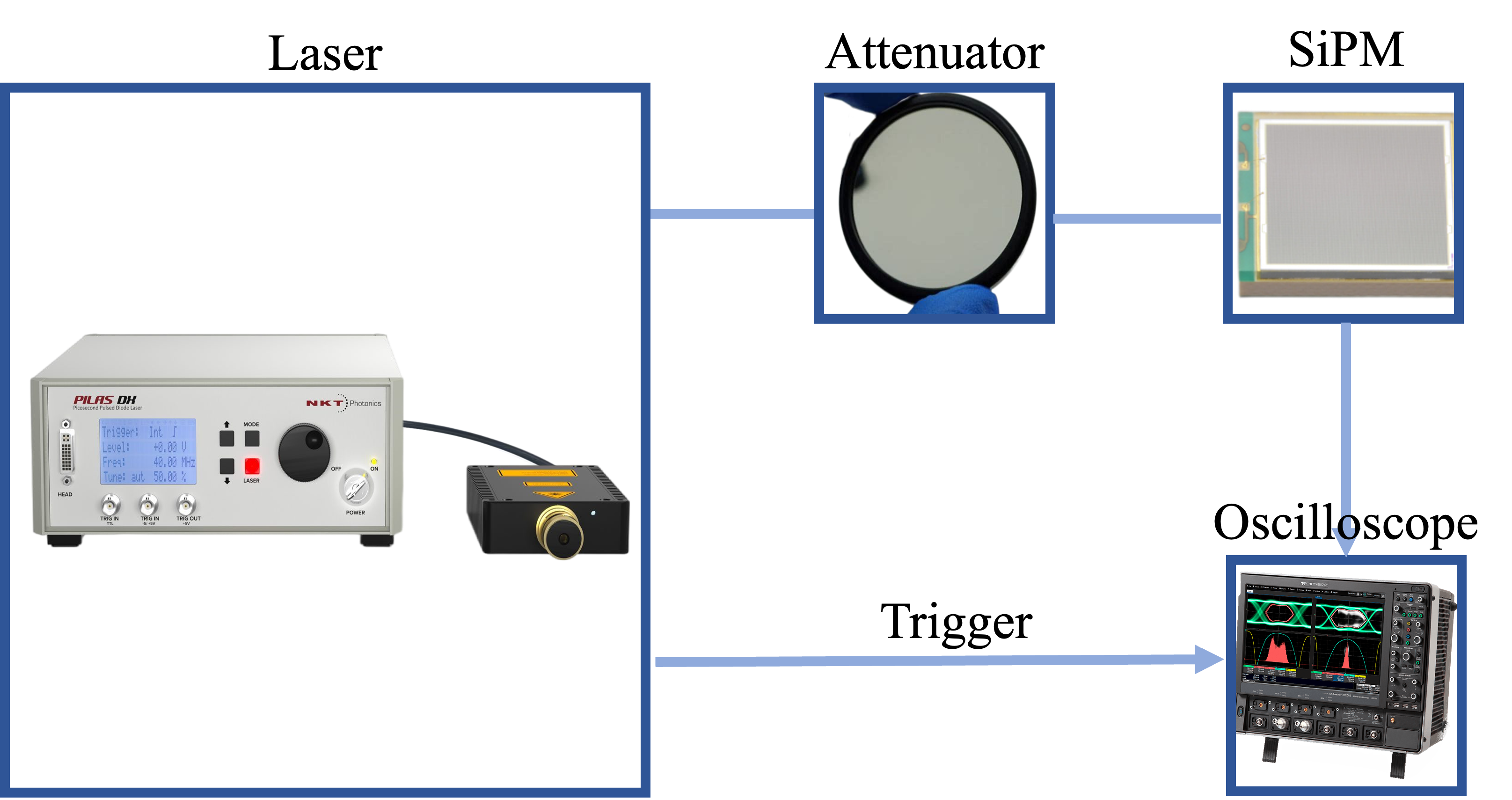}
\caption{The setup of the SiPM testing system (SiPM System), which is used to quantify the $N_\mathrm{p.e.}$ and time resolution of the SiPM signal.}
\label{fig:SiPM_System}
\end{figure}

\paragraph{Experimental setups} Fig.~\ref{fig:SiPM_PIST_System} shows the experimental setup to characterise the response of the PIST ASIC to the SiPM signal (a.k.a. SiPM+PIST system), which is composed of a pico-second laser diode (with a peak wavelength at $405~\mathrm{nm}$), a light attenuator (to adjust the laser intensity), a pair of active differential probes ($6~\mathrm{GHz}$ bandwidth), and a high-speed oscilloscope (16 GHz bandwidth, 40 GS/s sampling rate). The laser intensity is adjustable to quantify the impacts of the amplitude on the timing resolution and the ToT response linearity. As there is no digitisation part within the PIST chip, the PIST output signals were recorded by the oscilloscope and used later for offline data analysis. Additionally, to characterise the signals that the SiPMs input to the ASIC and to further decouple the intrinsic timing performance of the PIST ASIC response to the SiPM signals, a variant experimental setup was constructed without the PIST ASIC (a.k.a. SiPM system), as shown in Fig.~\ref{fig:SiPM_System}.

\paragraph{SiPM characterization} Two types of SiPMs were selected, including MPPC S13360-6025PE from Hamamtatsu Photonics (HPK) and SiPM EQR06 11-3030D-S from Novel Device Laboratory (NDL), in order to cover a maximum-possible dynamic range, e.g. stringently required by future electromagnetic calorimetry. The waveforms of the two SiPM can be described by an exponential function as below
\begin{equation}
    V_{t}= V_{A} \cdot (1-e^{\frac{-(t-\tau_{0})}{\tau_{1}}}) \cdot e^{\frac{-(t-\tau_{0})}{\tau_{2}}}
\label{equation_expo}
\end{equation}
where $V_{A}$ is correlated to the waveform amplitude, $\tau_{0}$ denotes the SiPM signal start time, and $\tau_{1}$ and $\tau_{2}$ are exponential constants in the leading edge and the trailing edge, respectively. The characteristics of the two SiPMs are summarized in Table~\ref{Table:SiPMs}. The SiPM signals have been calibrated into the number of detected photoelectrons ($N_\mathrm{p.e.}$) based on the single photon charge of each SiPM type, with a typical $N_\mathrm{p.e.}$ of 200 for the 1 MIP signal. 

\begin{table}
  \begin{center}
    \caption{Characteristics of the two selected SiPMs.}
    \label{Table:SiPMs}
    \begin{threeparttable}
    \begin{tabular}{ccc}
    \hline
    {Type} & {S13360-6025PE~\cite{S13360-6025PE}} & {EQR06 11-3030D-S~\cite{EQR06}}\\
    \hline
    {Pixel pitch $\mathrm{(\mu m^{2})}$} & {$25\times25$} & {$6\times6$}\\
    {Effective photosensitive area $\mathrm{(mm^{2})}$} & {$6\times6$} & {$3\times3$}\\
    {Number of pixels} & {57600} & {244720}\\
    {Ternimal capacitance $\mathrm{(pF)}$} & {1280} & {45.9}\\
    {Gain} & {$7\times10^{5}$} & {$8\times10^{4}$} \\
    {Waveform} & Fig.~\ref{fig:HPK_Waveform} & Fig.~\ref{fig:NDL_Waveform} \\
    {$\tau_{1}~\mathrm{(ns)}$} & 0.81 & 1.22 \\
    {$\tau_{2}~\mathrm{(ns)}$} & 62 & 4.22 \\
    \hline
    \end{tabular}
    \end{threeparttable}
  \end{center}
\end{table}

\begin{figure}[htbp]
\centering
\subfigure[A typical HPK-MPPC waveform corresponding to $N_\mathrm{p.e.}=3614.77$.]{
\begin{minipage}[c]{1\textwidth}
        \centering
        \includegraphics[width=1\textwidth]{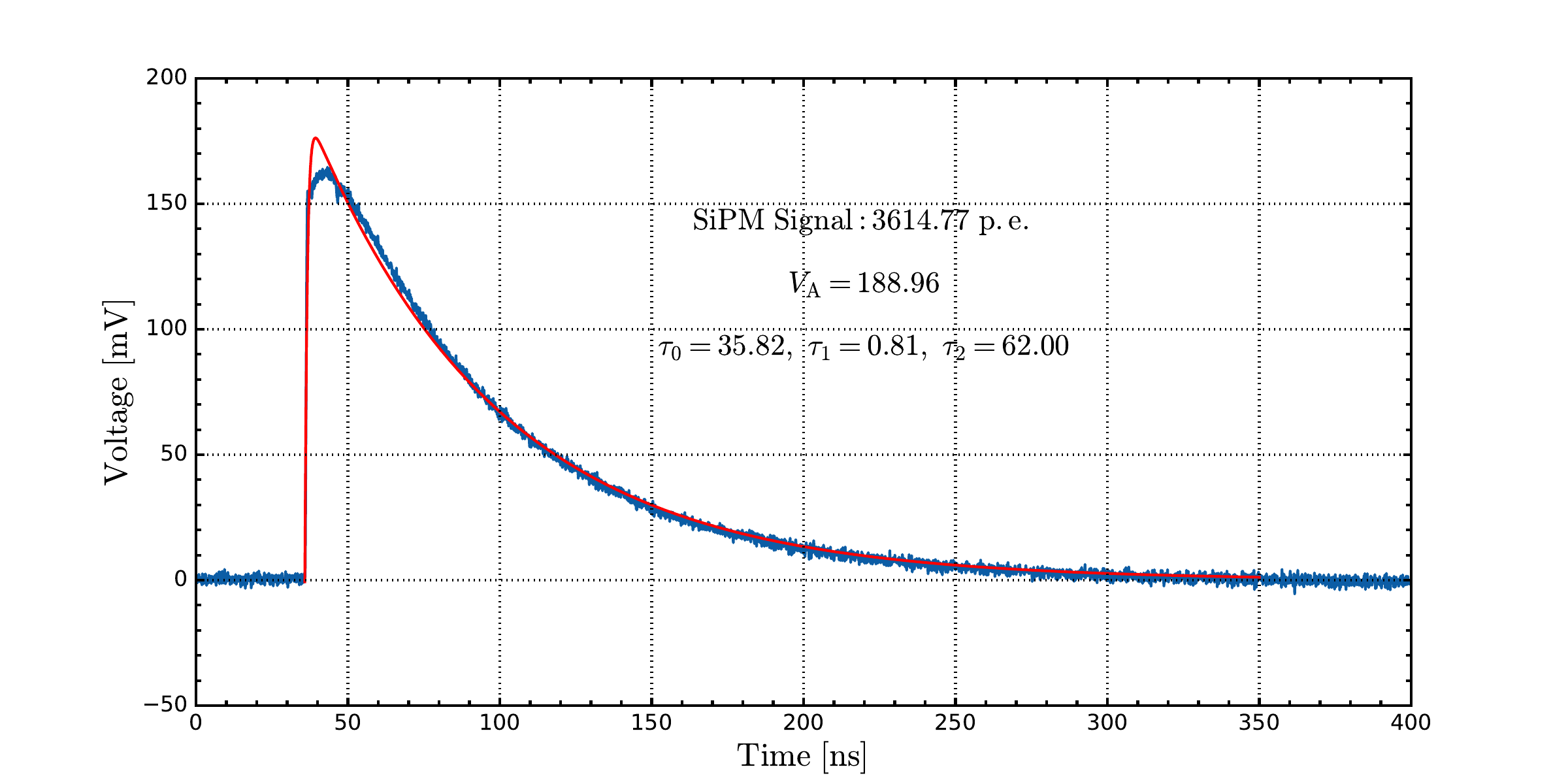}
        \label{fig:HPK_Waveform}
    \end{minipage}
}
\quad
\subfigure[A typical NDL-SiPM waveform corresponding to $N_\mathrm{p.e.}=3349.57$.]{
\begin{minipage}[c]{1\textwidth}
        \centering
        \includegraphics[width=1\textwidth]{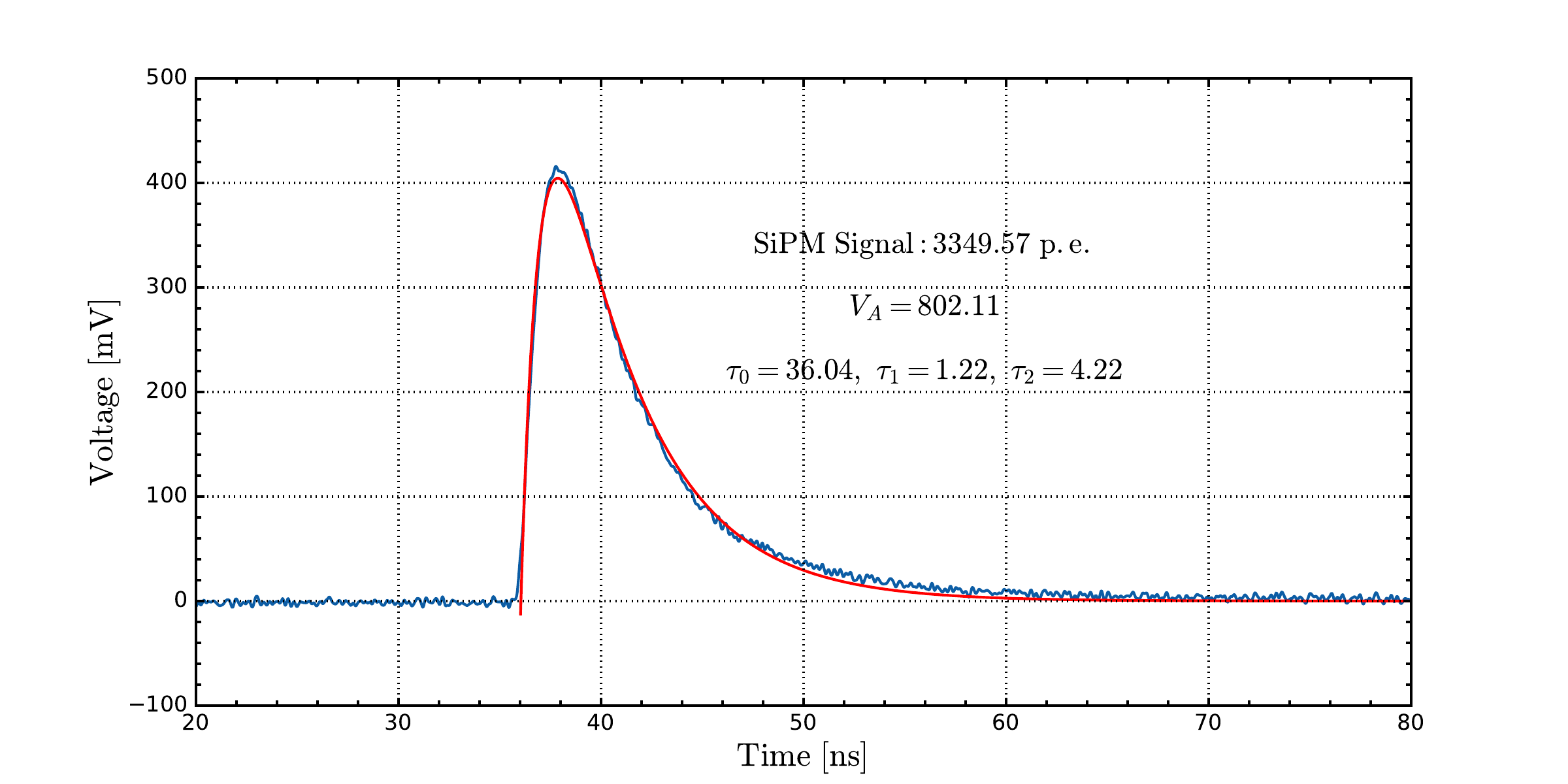}
        \label{fig:NDL_Waveform}
    \end{minipage}
}
\caption{The waveforms of the two SiPMs recorded by the oscilloscope.}
\label{fig:SiPM_Waveforms}
\end{figure}

\paragraph{PIST ASIC waveform description}

\begin{figure}
\centering
    \includegraphics[height=7cm]{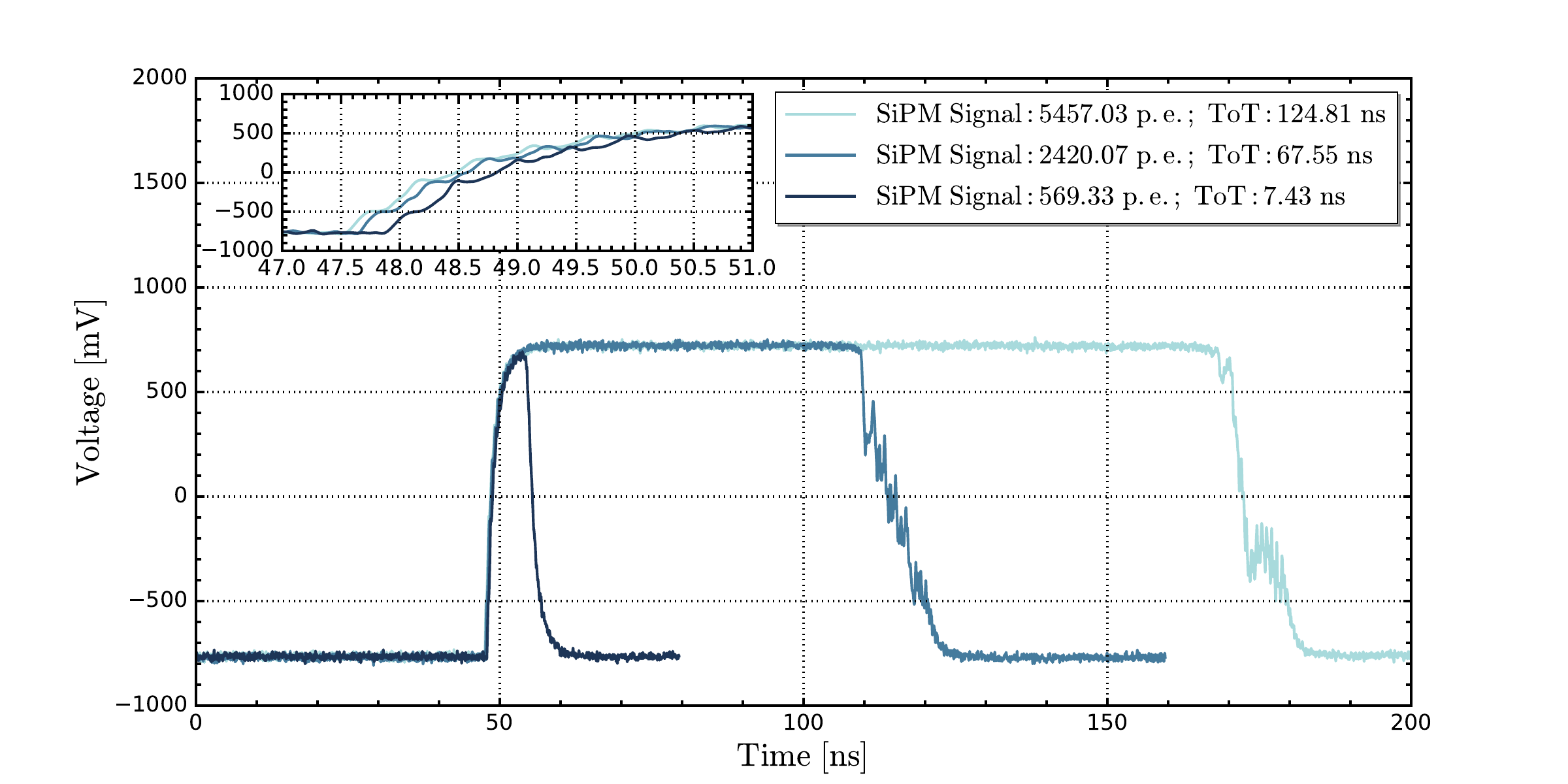}
\caption{The typical waveforms of the PIST ASIC vary with the $N_\mathrm{p.e.}$ of the SiPM signal. In the subplot, the leading edge of the waveform is magnified for display.}
\label{fig:ASIC_waveform}
\end{figure}

\begin{figure}
\centering
    \includegraphics[height=7cm]{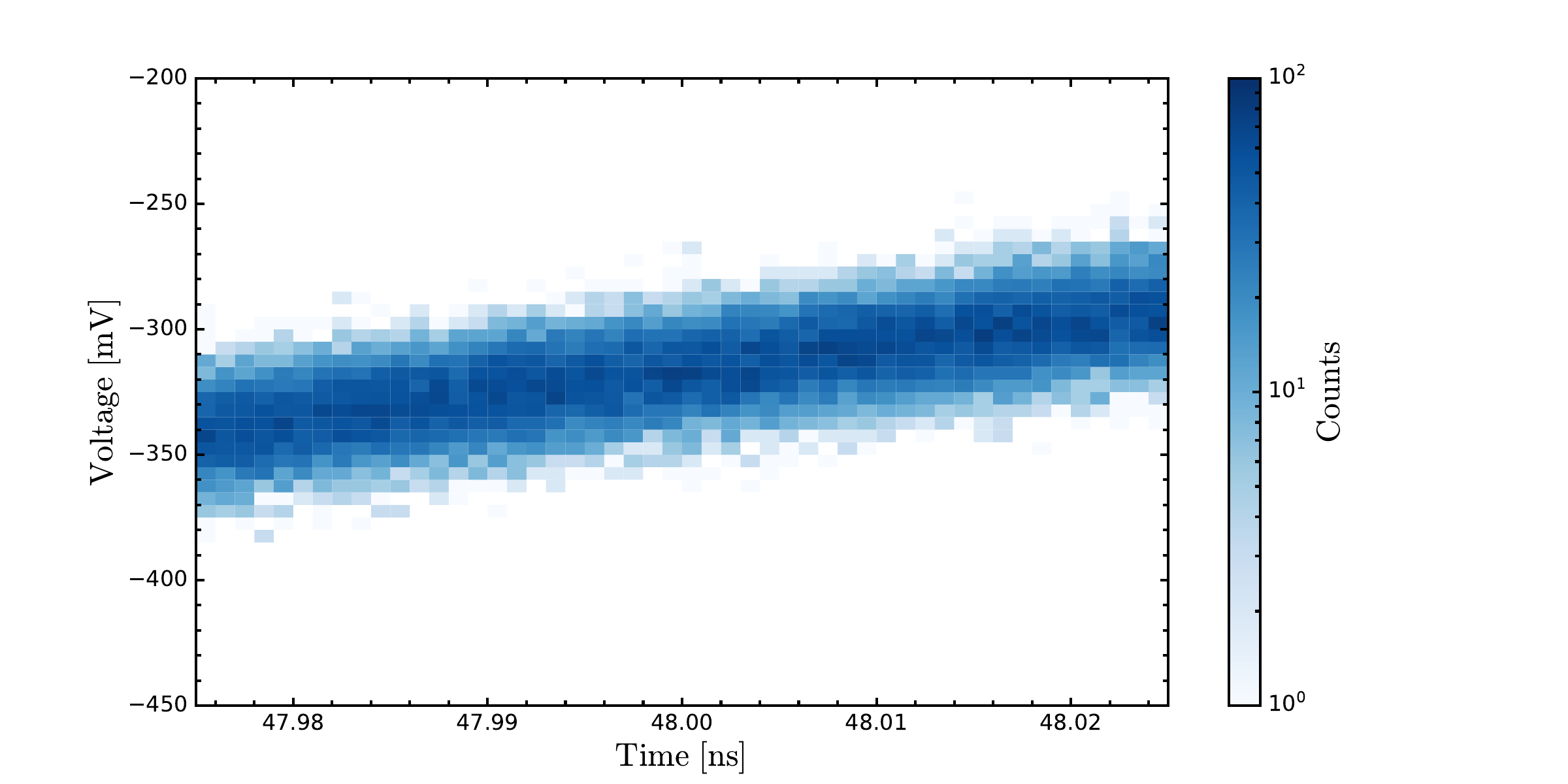}
\caption{Persistence display of the leading edge local area of the PIST ASIC waveforms.}
\label{fig:ASIC_local_waveform}
\end{figure}

The waveform of the ASIC signal is depicted in Fig.~\ref{fig:ASIC_waveform}. In single-ended pulse simulations, the voltage dynamic range of the PIST ASIC spans from $270~\mathrm{mV}$ to $1200~\mathrm{mV}$, corresponding to a differential equivalence of $-930~\mathrm{mV}$ to $930~\mathrm{mV}$~\cite{Lu:2023chl}. However, practical measurements reveal that the voltage dynamic range of the differential voltage is observed to be from $-780~\mathrm{mV}$ to $780~\mathrm{mV}$, indicating some gain compression, as shown in Fig.~\ref{fig:ASIC_waveform}, which can be attributed to the probe. Upon examination of the waveform, it is evident that the signal exhibits a smooth leading edge, which carries crucial timing information. Fig.~\ref{fig:ASIC_local_waveform} illustrates the persistence display of the leading edge derived from 10,000 events acquired at a sampling rate of $40~\mathrm{GS/s}$, thereby elucidating the noise characteristics associated with the leading edge.

\subsection{Analysis methodology}
\label{Sec:Ana}
\paragraph{Timing analysis method} To quantify the timing performance of the test stands, we developed an analysis method using the constant fraction discrimination (CFD)~\cite{Genat:2008vc} to obtain the time information of the signals. The time interval ($\delta t$) is defined as :
\begin{equation}
    \delta t = t_\mathrm{Leading} - t_\mathrm{Ref},
\end{equation}
where the $t_\mathrm{Leading}$ is defined as the time it takes for the leading edge to reach a fixed fraction of its maximum amplitude, the $t_\mathrm{Ref}$ is the reference time provided by the laser signal. For each laser intensity, we sampled 10,000 signal waveforms to obtain the distribution of the time intervals, and defined the time resolution ($\sigma$) as the standard deviation of the time interval distribution, which was fitted with a Gaussian function to represent the timing performance. By analyzing the $\mathrm{SiPM}+\mathrm{PIST}$ system, we can obtain $\sigma_\mathrm{SiPM+PIST}$, and by analyzing the SiPM system, we can obtain $\sigma_\mathrm{SiPM}$, which allows for the decoupling of the intrinsic timing performance of the PIST ASIC using the equation: 
\begin{equation}
\sigma_\mathrm{PIST} = \sqrt{\sigma_\mathrm{SiPM+PIST}^{2}-\sigma_\mathrm{SiPM}^{2}}.
\end{equation}

\begin{figure}[htbp]
\centering
\subfigure[]{
\begin{minipage}[c]{.45\textwidth}
        \centering
        \includegraphics[width=1\textwidth]{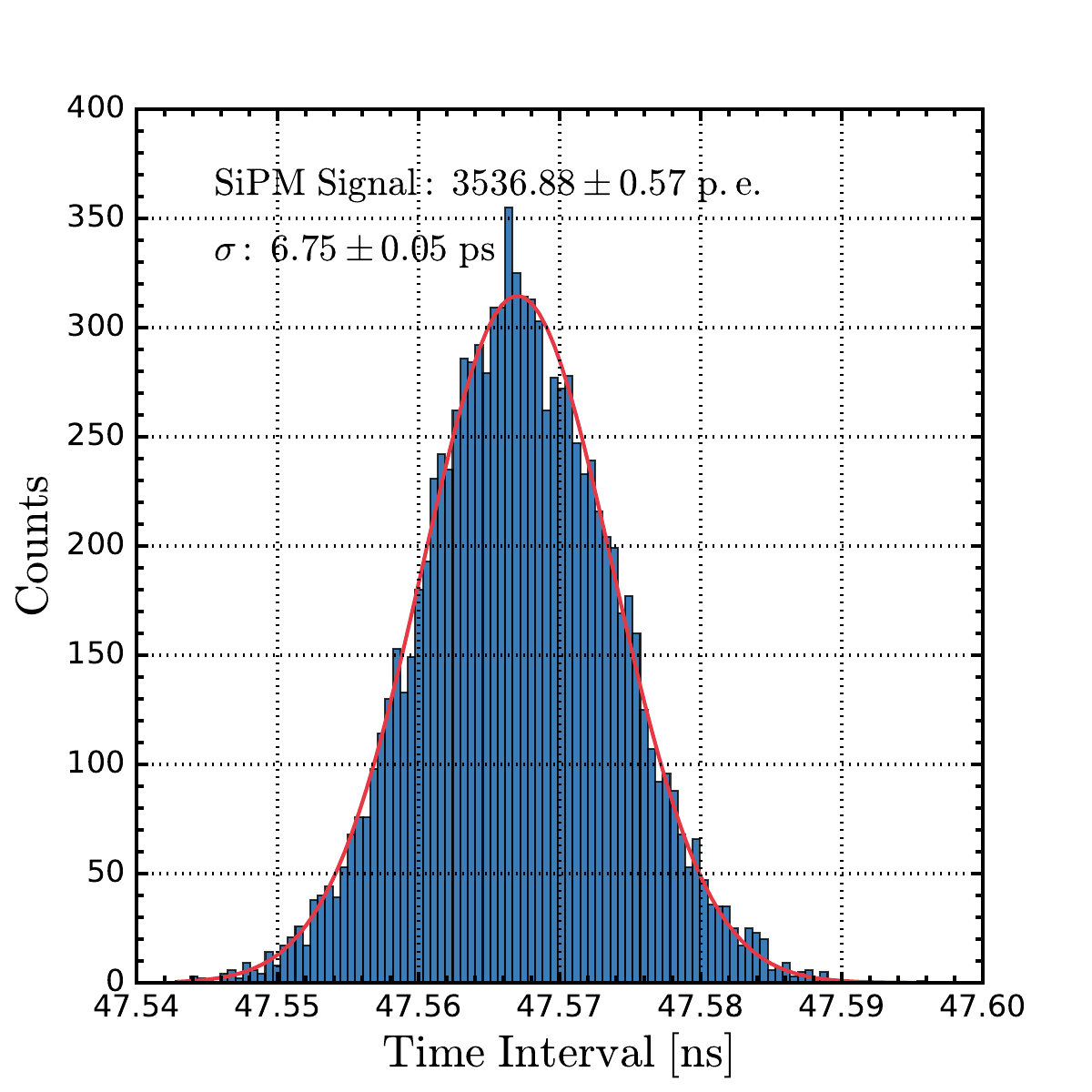}
        \label{fig:HPK_PE_histo}
    \end{minipage}
}
\quad
\subfigure[]{
\begin{minipage}[c]{.45\textwidth}
        \centering
        \includegraphics[width=1\textwidth]{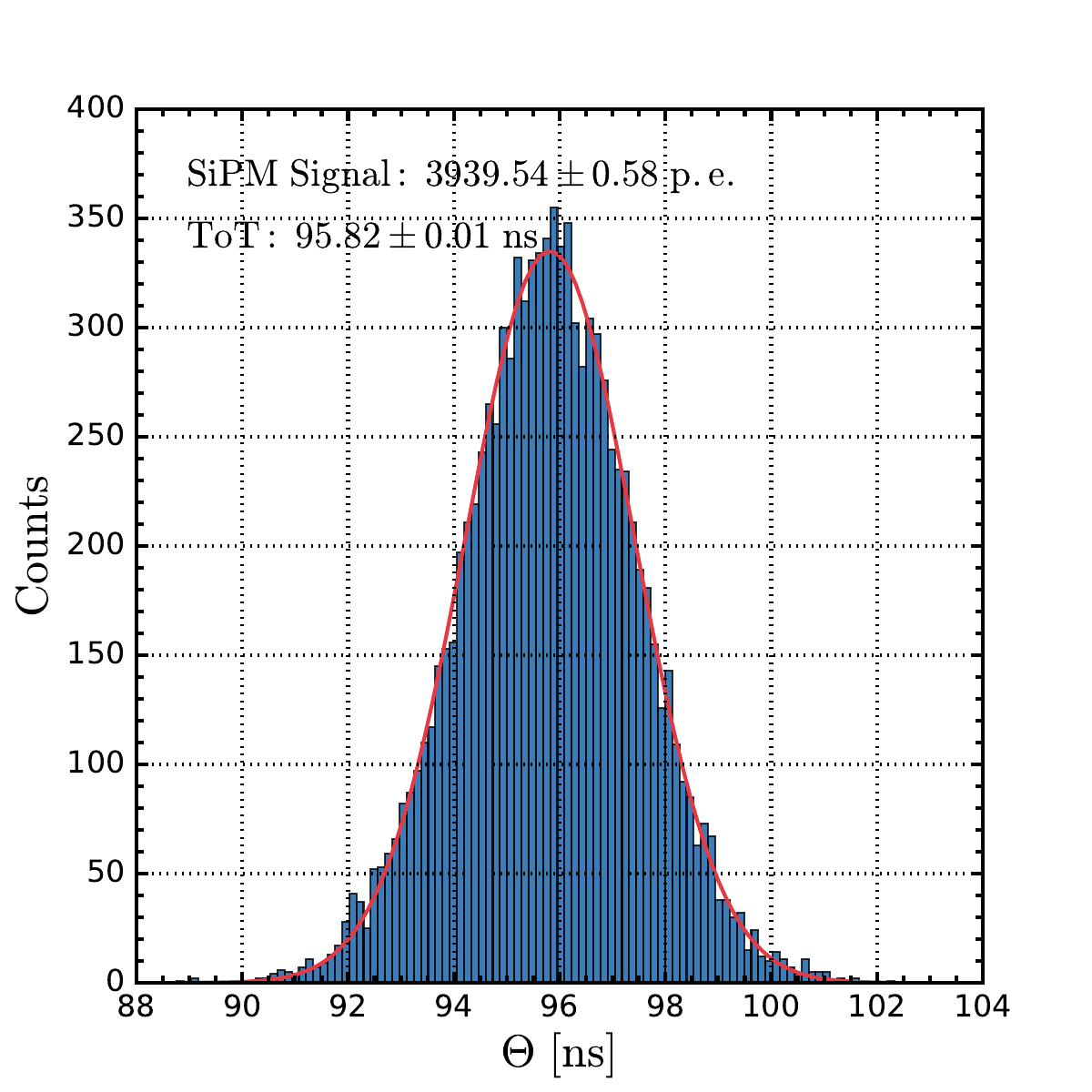}
        \label{fig:HPK_ToT_histo}
    \end{minipage}
}
\label{fig:distribution}
\caption{For a SiPM signal with $N_\mathrm{p.e.}=3536.88$, distributions of time intervals (Fig.~\ref{fig:HPK_PE_histo}) and $\Theta$ (Fig.~\ref{fig:HPK_ToT_histo}) of 10,000 ASIC waveforms are obtained, with $t_\mathrm{Leading}$ set at 10\% CFD and $t_\mathrm{Trailing}$ at 50\% CFD.}
\end{figure}

\paragraph{ToT analysis method} We established an analysis method utilizing CFD to quantify the ToT dynamic range of the test stands, similar to the timing method used for quantifying timing performance. The ToT of an ASIC signal is defined as: 
\begin{equation}
    \Theta = t_\mathrm{Trailng} - t_\mathrm{Leading},
\end{equation}
where the $t_\mathrm{Leading}$ and $t_\mathrm{Trailing}$ are defined as the times at which the ASIC signal reaches a certain fraction of its maximum amplitude in the leading and trailing edges, respectively. We repeated the ASIC waveform sampling 10,000 times for each light intensity, and defined the ToT for that intensity as the expect value obtained by fitting the distribution of $\Theta$ from these 10,000 events with a Gaussian.

\section{Experimental results of the SiPM+PIST system}
\label{sec:results}
In this section, experimental results will be presented, including timing performance of the SiPM-PIST setup and the detailed evaluation of the ToT readout.

\begin{figure}[htbp]
\centering

\subfigure[]{
\begin{minipage}[c]{1\textwidth}
        \centering
        \includegraphics[width=1\textwidth]{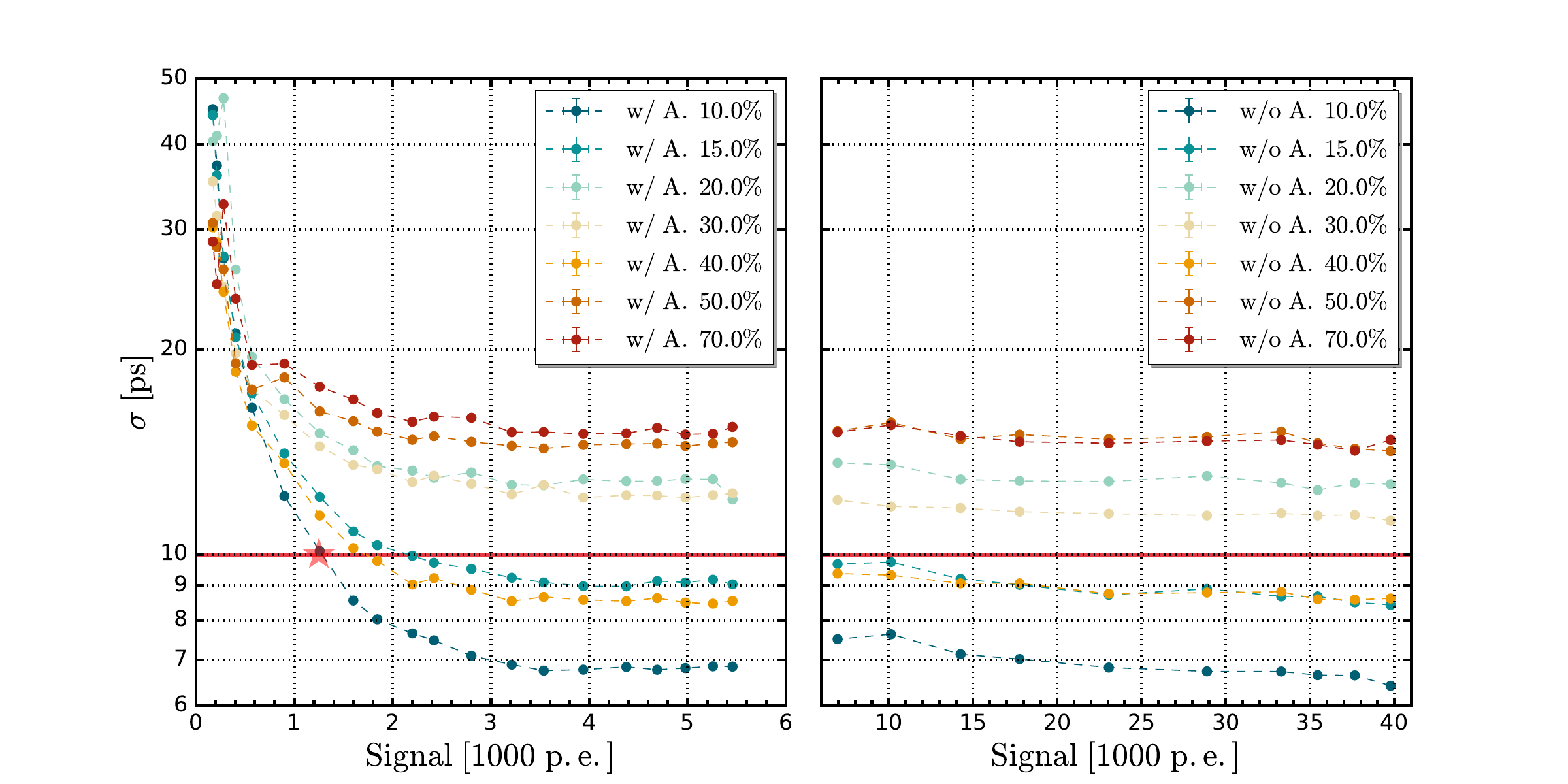}
    \label{fig:HPK_ASIC_TR}
    \end{minipage}
}
\quad
\subfigure[]{
\begin{minipage}[c]{1\textwidth}
        \centering
        \includegraphics[width=1\textwidth]{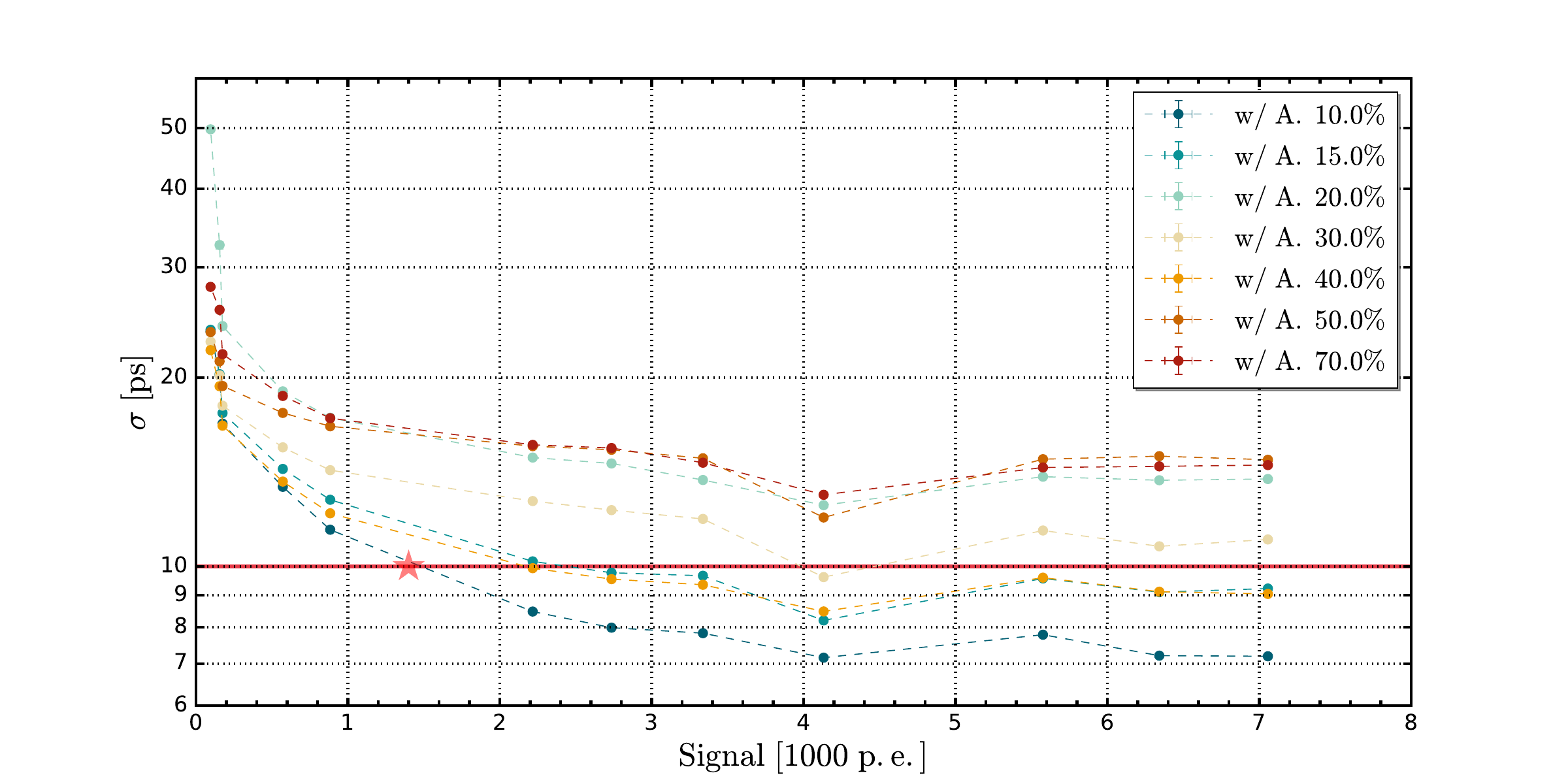}
    \label{fig:NDL_ASIC_TR}
    \end{minipage}
}
\caption{The time resolution of the SiPM-PIST system varies with the SiPM signal's $N_\mathrm{p.e.}$ (Fig.~\ref{fig:HPK_ASIC_TR}: HPK SiPM, Fig.~\ref{fig:NDL_ASIC_TR}: NDL SiPM) and the trigger threshold in CFD (full scan). Fig.~\ref{fig:HPK_ASIC_TR} includes two subplots, the left subplot with the attenuator and the right subplot without, for a larger dynamic range.}
\label{fig:Time_performance}
\end{figure}

\paragraph{Timing performance} The test results of the time resolution as a function of the $N_\mathrm{p.e.}$ detected by HPK MPPC and NDL SiPM, along with full scans of the CFD trigger threshold, are presented in Fig.~\ref{fig:Time_performance}. The test results indicate the following points: 
\begin{itemize}
\item [1)] The 10\% CFD threshold turns out to be optimal for the timing performance of PIST ASIC signals, as it offers a steeper slope, resulting in an enhanced time resolution without compromising due to noise. 
\item [2)] Generally, the timing resolution can be improved with larger SiPM signals and enters a plateau region when the SiPM signal reaches more than $3,000~\mathrm{p.e.}$.
\item [3)] The system timing resolution is below $50~\mathrm{ps}$ across the entire dynamic range of $170~\mathrm{p.e.}$ to $40,000~\mathrm{p.e.}$.
\item [4)] When using a 10\% CFD threshold with a $1~\mathrm{MIP}$ signal ($N_\mathrm{p.e.} = 200$), the time resolution of the test system is with the HPK SiPM input to the ASIC and the NDL SiPM is $45~\mathrm{ps}$ and $24~\mathrm{ps}$, respectively.
\item [5)] With a 10\% CFD threshold, the time resolution of the test system is better than $10~\mathrm{ps}$ when the signal $N_\mathrm{p.e.}$ exceeds 1200 and stays constant at around $7~\mathrm{ps}$ with larger signals.
\end{itemize}

\begin{figure}[htbp]
\centering
\subfigure[]{
\begin{minipage}[c]{.45\textwidth}
        \centering
        \includegraphics[width=1\textwidth]{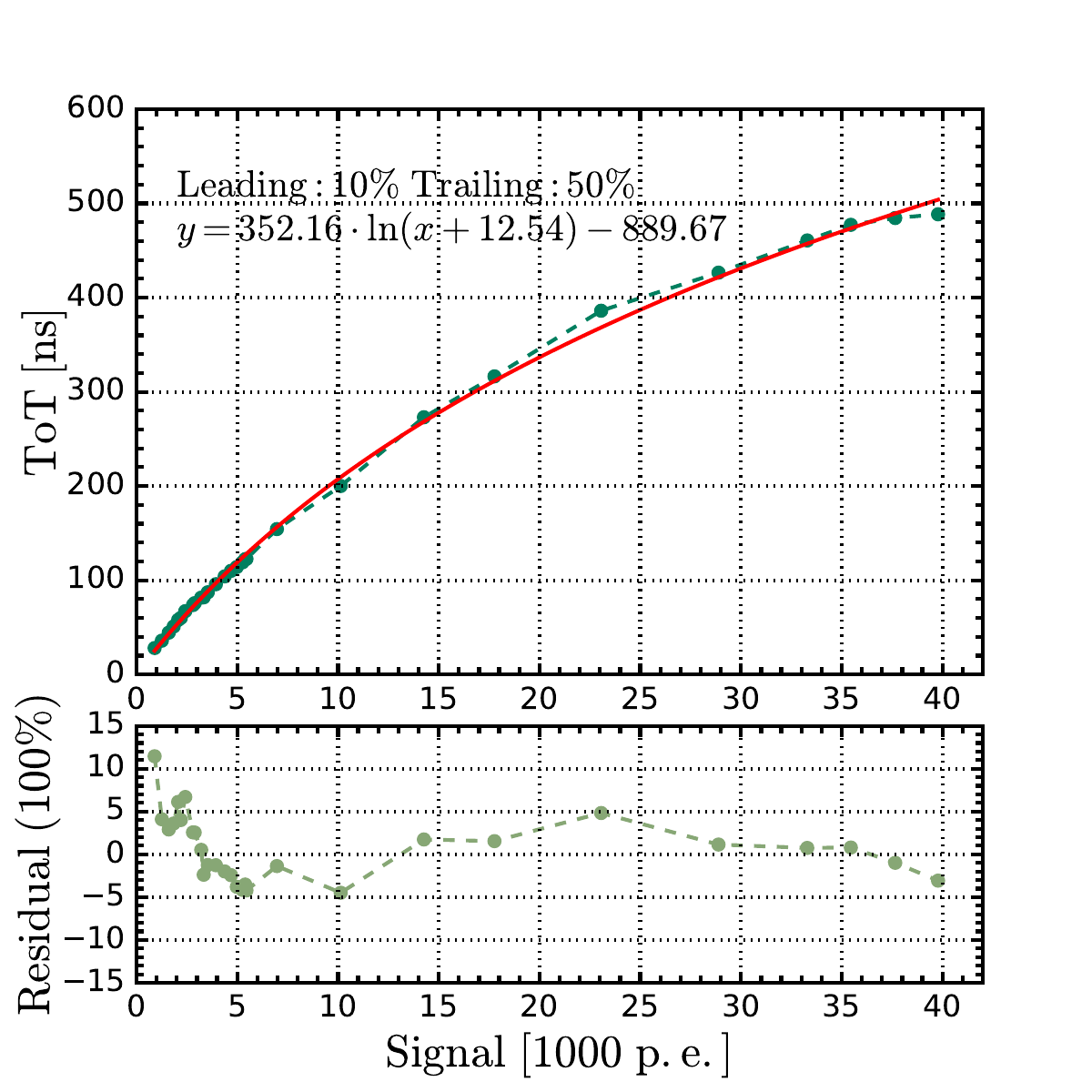}
    \label{fig:HPK_ASIC_TOT_log}
    \end{minipage}
}
\quad
\subfigure[]{
\begin{minipage}[c]{.45\textwidth}
        \centering
        \includegraphics[width=1\textwidth]{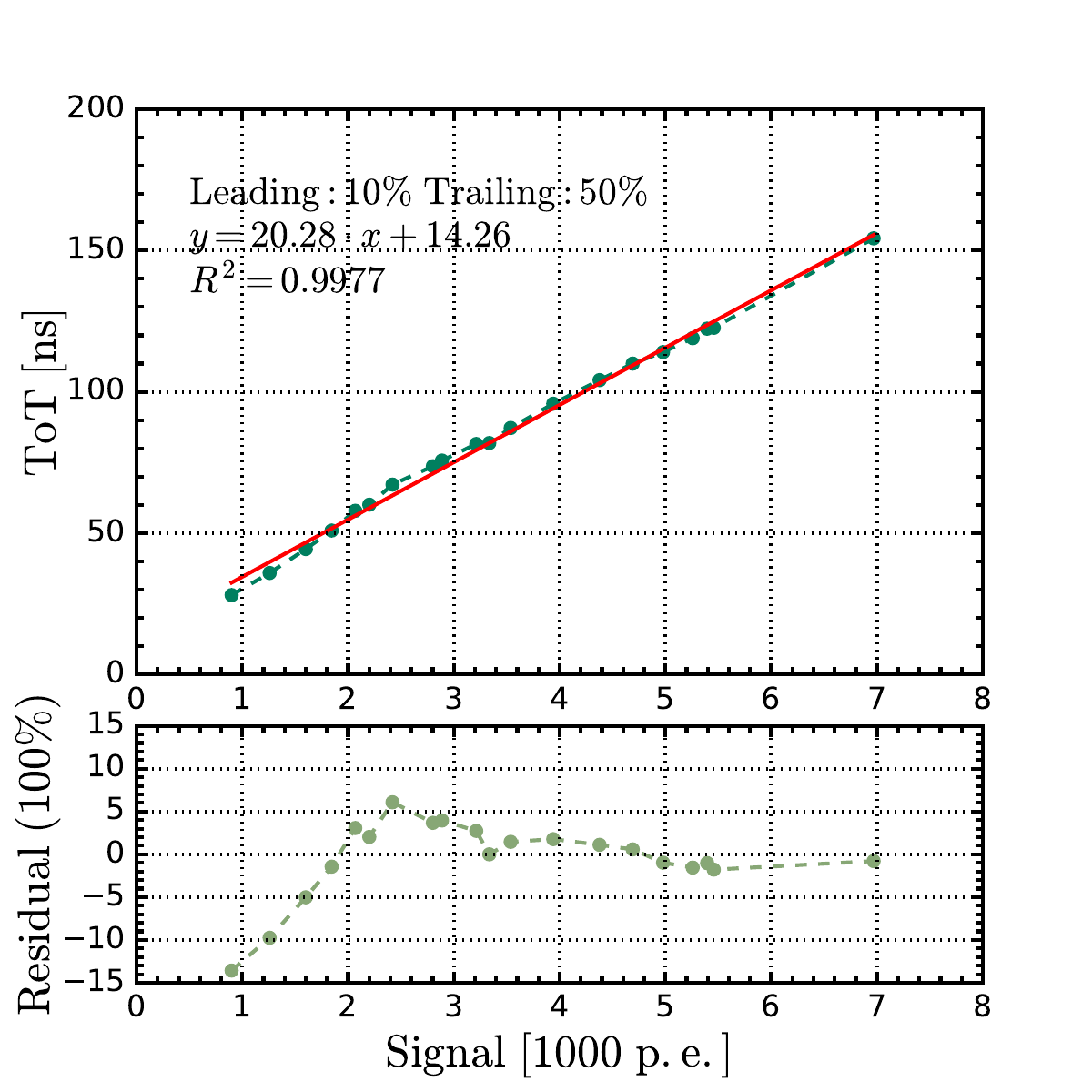}
    \label{fig:HPK_ASIC_TOT_line}
    \end{minipage}
}
\caption{The ToT varies with the HPK SiPM signal's $N_\mathrm{p.e.}$ ($t_\mathrm{Leading}$ set at 10\% CFD and $t_\mathrm{Trailing}$ at 50\% CFD). In Fig.~\ref{fig:HPK_ASIC_TOT_log}, within the range from $899.02~\mathrm{p.e.}$ to $39780.02~\mathrm{p.e.}$, the relationship fits a logarithmic equation: $ToT = 352.16 \cdot ln(N_\mathrm{p.e.} + 12.54) - 889.67$. In Fig.~\ref{fig:HPK_ASIC_TOT_line}, in the range of $899.02~\mathrm{p.e.}$ to $6967.86~\mathrm{p.e.}$, the relationship fits a linear equation: $ToT = 20.28 \cdot N_\mathrm{p.e.} + 14.26$, with $R^{2} = 0.9977$. The residuals between the equation and experimental results are within 15\%.}
\end{figure}

\paragraph{Time over threshold} As shown in Fig.~\ref{fig:ASIC_waveform}, the amplitude of the ASIC signal remains constant regardless of the SiPM signal, while the ToT of the ASIC signal varies with the $N_\mathrm{{p.e.}}$. The definition of ToT can be found in Sec.~\ref{Sec:Ana}. Considering the smooth leading edge of the ASIC signal and its good timing performance at the lower end, contrasted with the significant jitter observed in the trailing edge (as seen in Fig.~\ref{fig:ASIC_waveform}), we opt to use different thresholds to define ToT for each edge. The leading edge threshold is defined as 10\% or 30\% of the amplitude, while the trailing edge threshold is set at 50\% of the amplitude. The measured output waveforms for signals of $569.33~\mathrm{p.e.}$, $2420.07~\mathrm{p.e.}$, and $5457.03~\mathrm{p.e.}$ are depicted in Fig.~\ref{fig:ASIC_waveform}, with corresponding ToT values defined at 10\% amplitude of the leading edge as $7.43~\mathrm{ns}$, $67.55~\mathrm{ns}$, and $124.81~\mathrm{ns}$, respectively. Additionally, for the $569.33~\mathrm{p.e.}$ signal, the rise time thresholds at 10\% and 30\% amplitude are associated with ToT values of $7.43~\mathrm{ns}$ and $7.06~\mathrm{ns}$, respectively. The ToT in the dynamic range of $899.02~\mathrm{p.e.}$ to $39780.02~\mathrm{p.e.}$ signal is quantified and illustrated in Fig.~\ref{fig:HPK_ASIC_TOT_log}, and the relation of ToT versus $N_\mathrm{{p.e.}}$ follows a logarithmic function~\cite{ORITA2018303}, which can be described as below:
\begin{equation}
ToT = p_1 \cdot \ln(N_\mathrm{p.e.}+N_0) + p_0 ,
\label{Eq:ToT}
\end{equation}
where $N_0$ is the offset of the input SiPM signal, $p_1 \cdot N_0 + p_0$ is the offset of the ToT response. For the SiPM+PIST system (with HPK SiPM), experimental test results were fitted using the Eq.~\ref{Eq:ToT}, resulting in $p_0=\mathrm{352.16~ns}$, $p_1=\mathrm{12.54~p.e.}$, $p_2=\mathrm{-889.67~ns}$, with residuals within a range of 15\%. In the dynamic range spanning from $899.02~\mathrm{p.e.}$ to $6967.86~\mathrm{p.e.}$ signal, the ToT demonstrates a strong linear relationship with $N_\mathrm{p.e.}$, exhibiting a linearity of 0.9977, described by the equation:
\begin{equation}
    ToT =  p_1\cdot N_\mathrm{p.e.} + p_0, 
\end{equation}
where $p_1=20.28$ ns and $p_0=14.26$ ns, as shown in Fig.~\ref{fig:HPK_ASIC_TOT_line}. The corresponding residuals are also within 15\%.

\begin{figure}
\centering
    \includegraphics[width=1\textwidth]{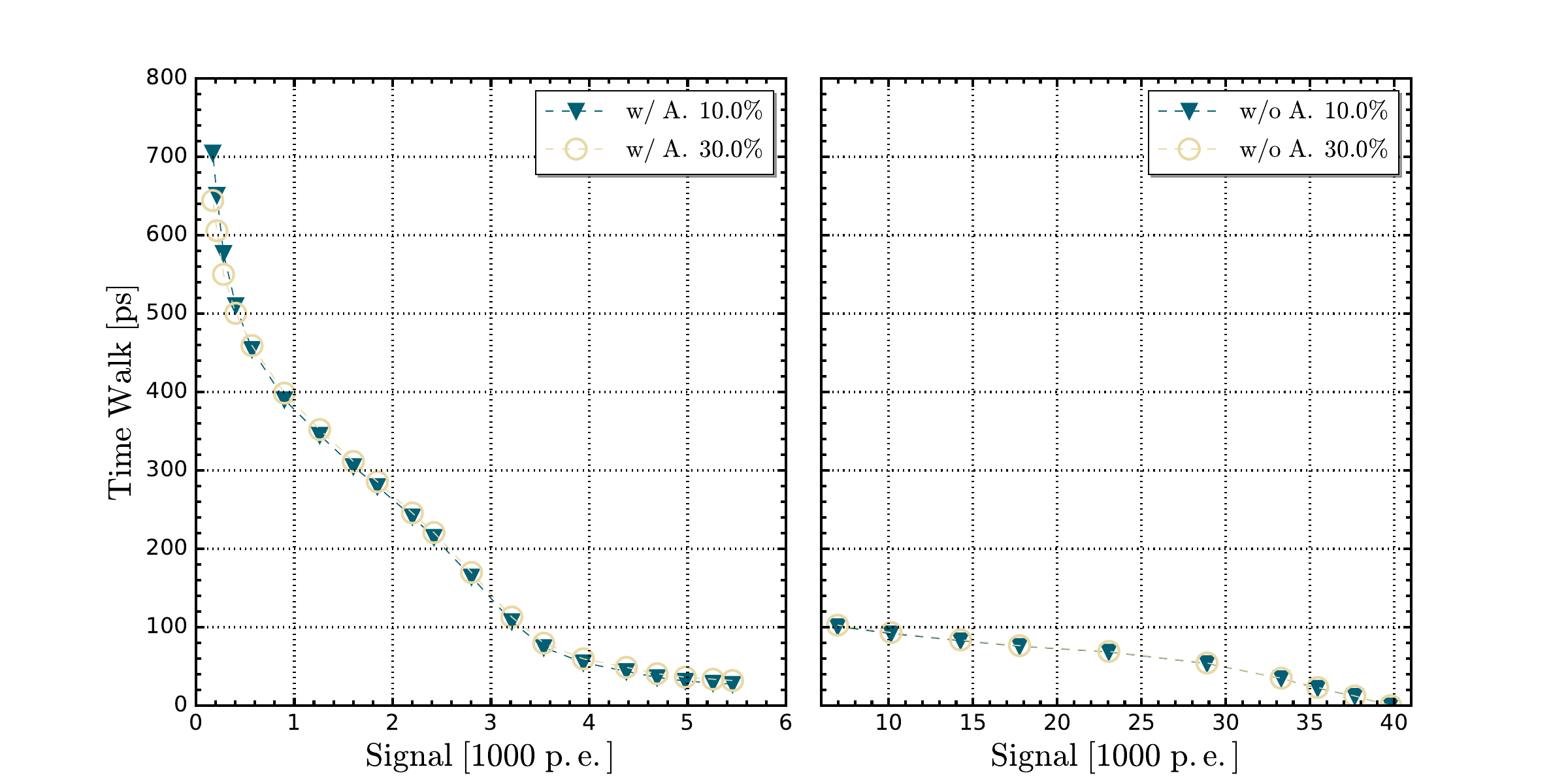}
\caption{The time walk of the PIST ASIC varies with the HPK SiPM signal's $N_\mathrm{p.e.}$ ($t_\mathrm{Leading}$ set at 10\% and 30\% CFD).}
\label{fig:HPK_TW}
\end{figure}

\paragraph{Time walk} In the subplot of Fig.~\ref{fig:ASIC_waveform}, as the $N_\mathrm{{p.e.}}$ increases, the signal's rise time decreases and the slope of the leading edge increases. Consequently, larger signals cross the threshold earlier than smaller ones, causing a reduction in the time interval. This effect, known as time walk, can affect the ToT response. To quantify the effect, time walk is defined as the relative difference between the threshold crossing time at different signal levels and that at the maximum signal. The relationship between time walk and the signal is shown in Fig.~\ref{fig:HPK_TW}, indicating a decrease in the time walk function with increasing $N_\mathrm{p.e.}$. Upon comparing the measured performances of ToT and time walk, the latter is relatively small, and any negative impacts attributed to time walk can be considered negligible. Additionally, no significant difference in the time walk effect is observed between 10\% CFD and 30\% CFD.

\section{Discussions}
\label{sec:discussion}

\begin{figure}[htbp]
\centering
\subfigure[]{
\begin{minipage}[c]{1\textwidth}
        \centering
        \includegraphics[width=1\textwidth]{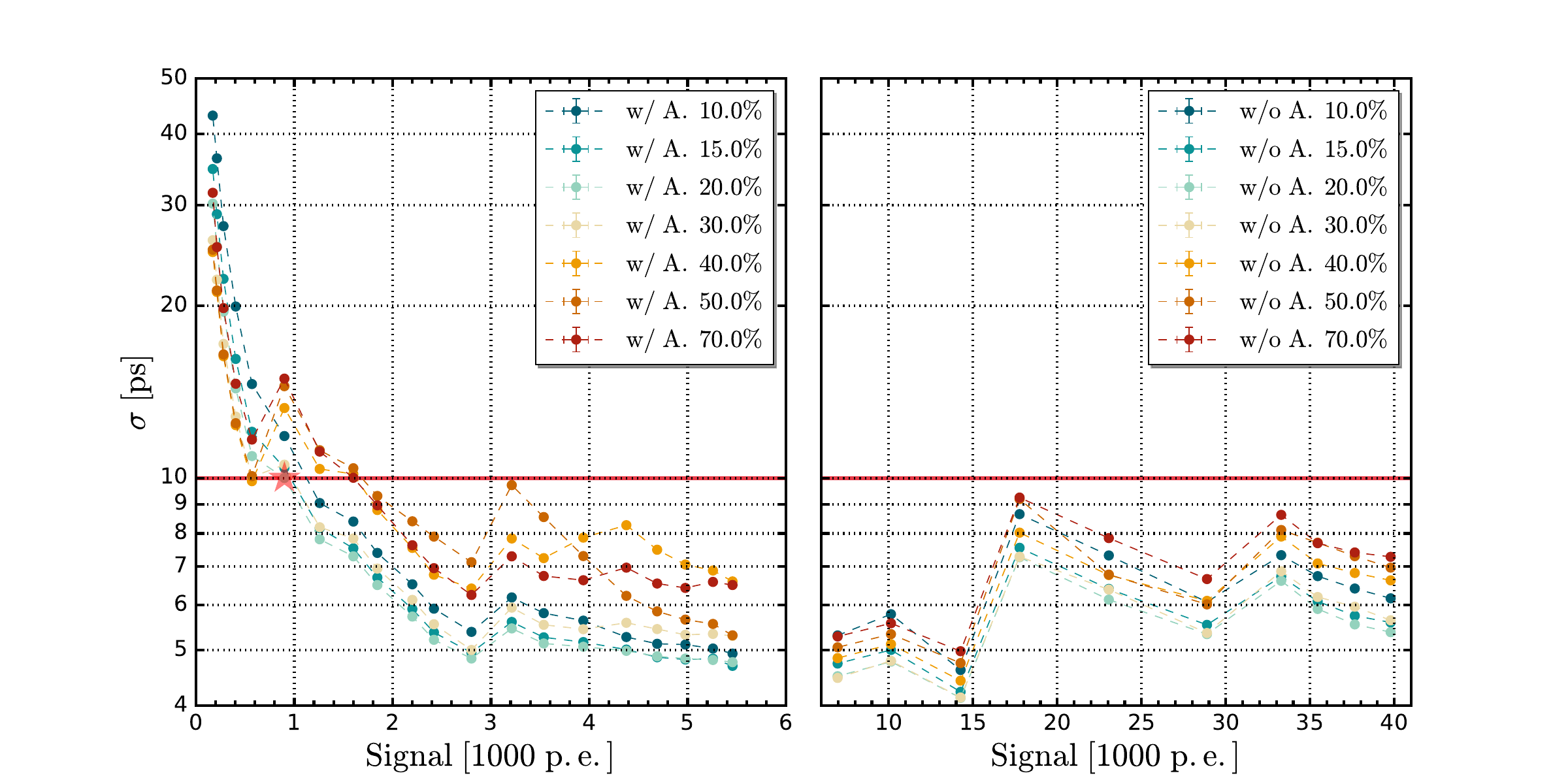}
        \label{fig:HPK_TR}
    \end{minipage}
}
\quad
\subfigure[]{
\begin{minipage}[c]{1\textwidth}
        \centering
        \includegraphics[width=1\textwidth]{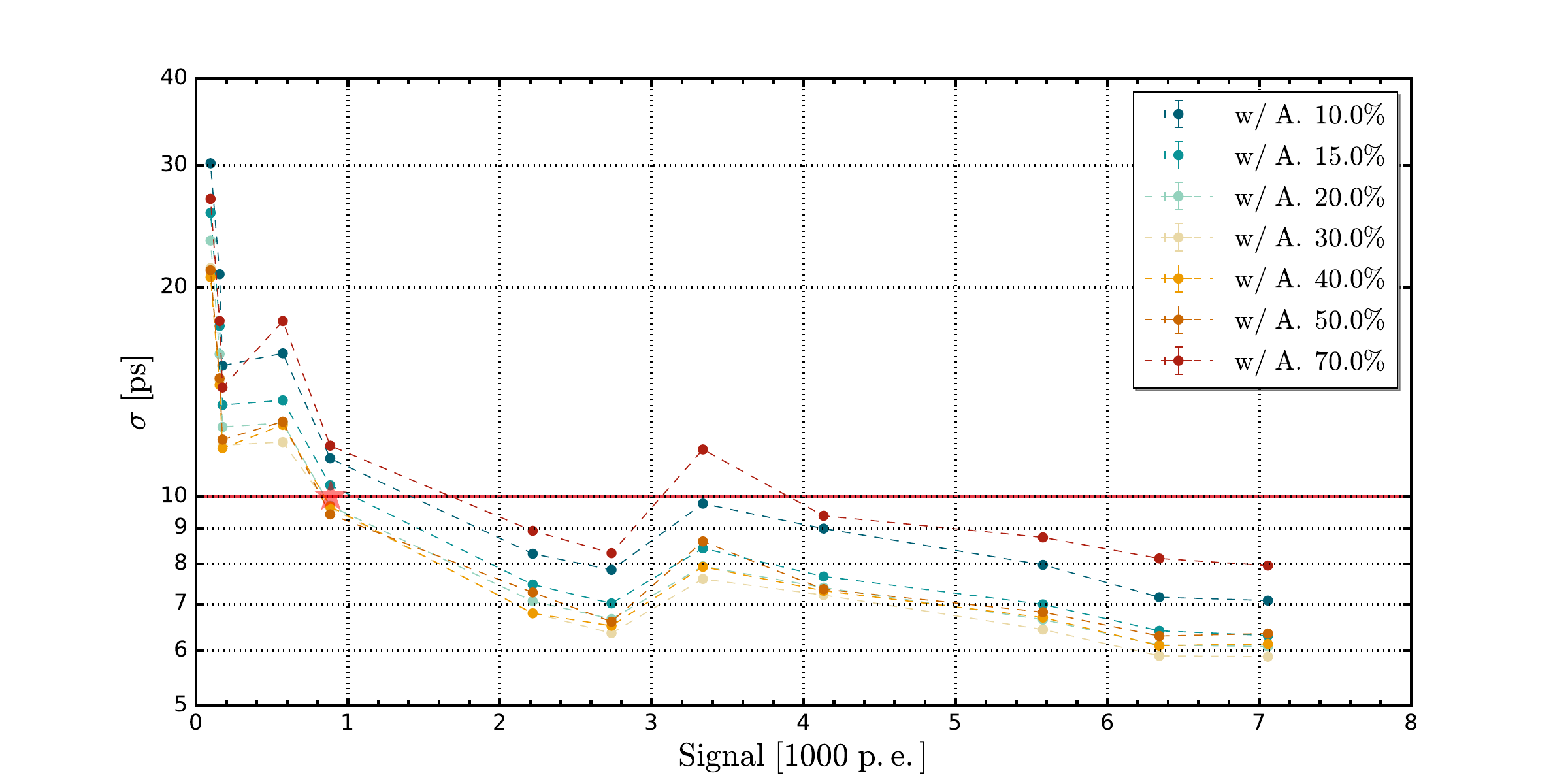}
        \label{fig:NDL_TR}
    \end{minipage}
}
\caption{The time resolution of the SiPM system varies with the SiPM signal's $N_\mathrm{p.e.}$ (Fig.~\ref{fig:HPK_TR}: HPK SiPM, Fig.~\ref{fig:NDL_TR}: NDL SiPM) and the trigger threshold in CFD (full scan). Fig.~\ref{fig:HPK_TR} includes two subplots, the left subplot with the attenuator and the right subplot without, for a larger dynamic range.}
\label{fig:SiPM_Sys_TR}
\end{figure}

\begin{figure}
\centering
    \includegraphics[width=0.5\textwidth]{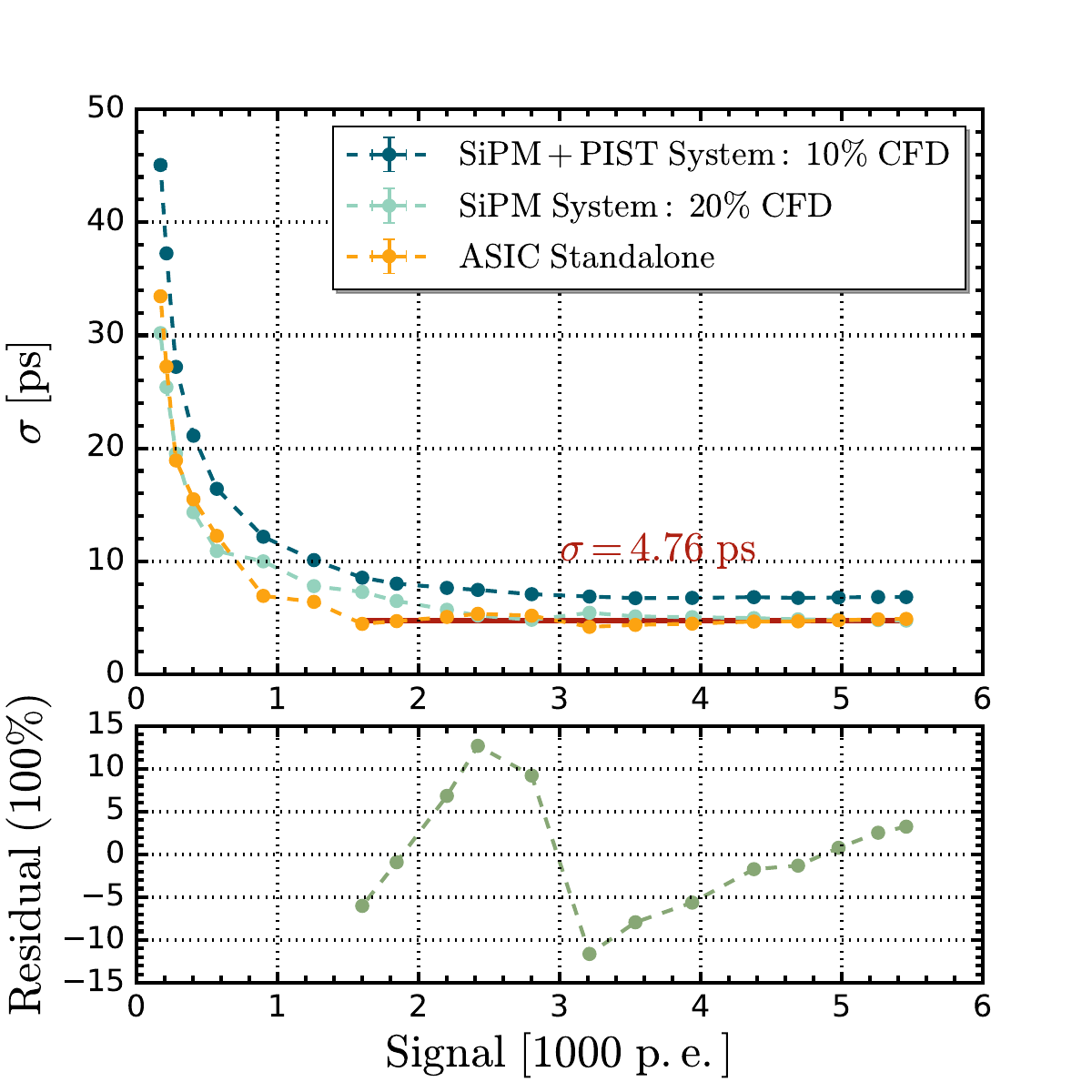}
\caption{The time resolution of the SiPM+PIST system (10\% CFD), SiPM system (20\% CFD) and PIST ASIC standalone varies with the HPK SiPM signal's $N_\mathrm{p.e.}$. The intrinsic time resolution of the PIST ASIC is calculated by $\sigma_\mathrm{PIST} = \sqrt{\sigma_\mathrm{SiPM+PIST,10\%}^{2}-\sigma_\mathrm{SiPM,20\%}^{2}}$. From $N_\mathrm{p.e.} = 1598.96$ to $N_\mathrm{p.e.} = 5458.07$, the time resolution of the PIST ASIC standalone fits the equation $\sigma_\mathrm{PIST} = 4.76$ with residual in 15\%.}
\label{fig:ASIC_standalone}
\end{figure}

\paragraph{Decomposition analysis of time resolution contributions} The time resolution of the SiPM+PIST system (Fig.~\ref{fig:SiPM_PIST_System}) presented in Sec.~\ref{sec:results} includes contributions from the laser, the SiPM, the PIST ASIC, signal cables, and the oscilloscope. Using the SiPM alone system with the laser (Fig.~\ref{fig:SiPM_System}), we can measure all other contributions to the time resolution except the PIST ASIC. Fig.~\ref{fig:SiPM_Sys_TR} shows the relation of time resolution versus the $N_\mathrm{p.e.}$ in the SiPM system. The trend (better time resolution with larger $N_\mathrm{p.e.}$) is consistent with the SiPM+PIST system, but there are some different points listed as below.
\begin{itemize}
    \item [1)] Comparing the timing performance at different thresholds, it turns out that the 20\% CFD threshold provides the best time resolution for the SiPM system, while the optimal value for the ASIC output signals is the 10\% CFD threshold due to significantly different waveform shapes of SiPM and PIST signals; 
    \item [2)] The distributions in Fig.~\ref{fig:SiPM_Sys_TR} exhibit a pronounced offset in the plateau region, whereas Fig.~\ref{fig:Time_performance} remains relatively stable. As the SiPM amplitude increases along with higher light intensity, the oscilloscope's voltage range was adjusted accordingly to avoid any signal saturation. However, the oscilloscope vertical resolution is fixed at 8-bit, thus changing the range will dramatically affect the time resolution. On the other hand, the PIST-ASIC amplitude is independent of the light intensity and is not affected by this factor.
\end{itemize}
The PIST-ASIC intrinsic time resolution can be estimated by the following equation, 
\begin{equation}
    \sigma_\mathrm{PIST} = \sqrt{\sigma_\mathrm{SiPM+PIST,10\%}^{2}-\sigma_\mathrm{SiPM,20\%}^{2}},
\end{equation}
and the result is shown in Fig.~\ref{fig:ASIC_standalone}, based on the assumption that the two systems are independent from each other. For the two systems, we choose different thresholds for the best time resolution of each system. Taking into account the impact of the oscilloscope's fixed voltage precision on the time resolution of the SiPM system, a fitting of the time resolutions in the plateau region (from $N_\mathrm{p.e.} = 1598.96$ to $N_\mathrm{p.e.} = 5458.07$) is conducted using the equation $\sigma_\mathrm{PIST} = a$ to determine the intrinsic timing performance of the ASIC. The time resolution obtained in the plateau region is $4.76\pm0.60~\mathrm{ps}$, which is close to the previous results of $4.2 \pm 0.04~\mathrm{ps}$~\cite{Lu:2023chl}. It is worth mentioning that we used a $1.5~\mathrm{m}$ long cable with the impedance of $50~\mathrm{ohm}$ and the bandwidth of 4~GHz to transmit the SiPM signal to the PIST-ASIC, and a pair of active differential probes with the bandwidth of 6~GHz to capture the ASIC output signal and convey it to the oscilloscope. However, during the sampling of the SiPM signal, we utilized a $1~\mathrm{m}$ long, cable with the impedance of $50~\mathrm{ohm}$ and the bandwidth of $4~\mathrm{GHz}$. The difference in the cable bandwidth is expected to affect the system's time resolution and introduce some systematic uncertainty in estimating the PIST intrinsic time resolution.

\begin{figure}
\centering
    \includegraphics[width=1\textwidth]{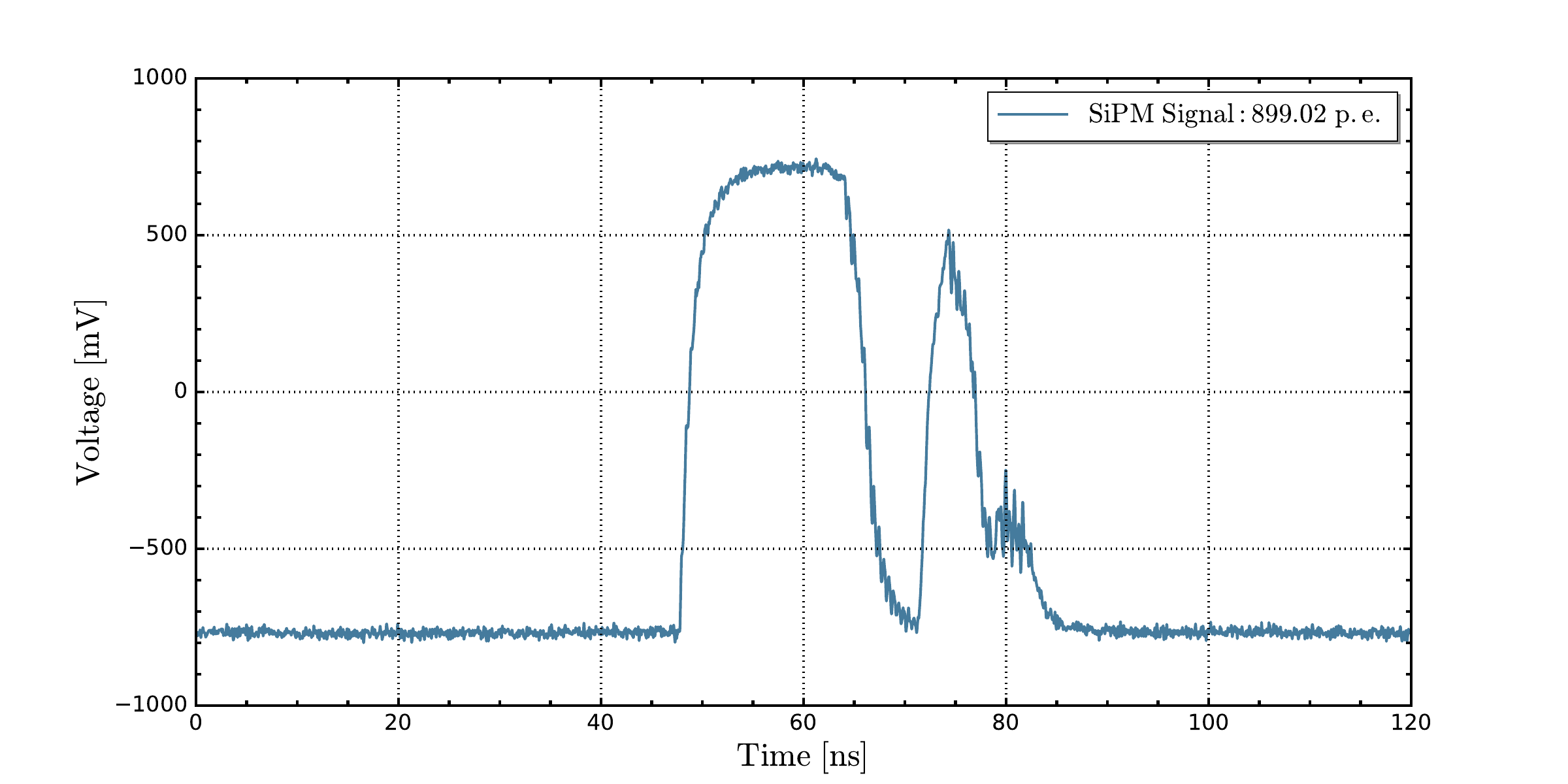}
\caption{A typical waveform of the PIST ASIC connected to the NDL-SiPM: the ToT response with several peak structures is significantly different from the nominal PIST output waveform (as shown in Fig.~\ref{fig:ASIC_waveform} ).}
\label{fig:ASIC_waveform_error}
\end{figure}

\paragraph{ToT response}
When analyzing the PIST-ASIC output signals and the input signals from the NDL SiPM, we observed oscillations in some signal waveforms (as Fig.~\ref{fig:ASIC_waveform_error} shows) and thus conjectured that oscillations are due to the trailing edge of the NDL-SiPM signals. Therefore, this issue would impact the ToT measurements, while the time resolution determined only by the signal rising edge remains unaffected.

\paragraph{SiPM+ASIC system versus others}

\begin{table*}
\begin{center}
\caption{Summary and comparison of fast timing ASICs performance.}
\label{Table:summary}
\begin{threeparttable}
    \resizebox{\linewidth}{!}{
    \begin{tabular}{ccccc}
    \hline
    \textbf{Parameter} & \multicolumn{4}{c}{Value} \\
    \hline
    \multirow{2}{*}{ASIC} & \multirow{2}{*}{PIST} & \multicolumn{2}{c}{CMS MTD} & ATLAS HGTD \\
    \cline{3-5}
    & & TOFHIR (BTL)~\cite{CMSMTD:2021imi}~\cite{Albuquerque:2020uyy}  & ETROC (ETL)~\cite{Sun:2021nlq} & ALTIROC1~\cite{Agapopoulou:2023jsd}  \\
    \hline
    CMOS Technology (nm)    & 55   & 130    & 65   & 130 \\
    Readout for             & SiPM  & SiPM  & LGAD  & LGAD \\
    Chip Area ($\mathrm{mm}$) & $3\times3$ & $8.5\times5.2$  & $20.8\times20.8$& $7.6\times7.7$ \\
    Intrinsic time resolution (ps) & $4.76~@~1600~\mathrm{p.e.}$ & $20~@~1~\mathrm{MIP}$ & $16~@~15~\mathrm{fC}$ & $15\pm1~@~10~\mathrm{fC}$ \\
    System time resolution (ps) & $7~@~1600~\mathrm{p.e.}$ & $22~@~4.2~\mathrm{MeV}$ & $29~@~30~\mathrm{fC}~\tnote{*}$ & $46.3\pm0.7$\\
    ToT Range & $900 \sim 40000~\mathrm{p.e.}$ & NA & $0 \sim 35~\mathrm{fC}$ & $0 \sim 40~\mathrm{fC}$ \\
    Power Dissipation (mW/Channel) & 15~\cite{Lu:2023chl} & $ 15~\tnote{**} $ & 404.48/606.27& 33.45 \\
    \hline
    \end{tabular}}
    
    \begin{tablenotes}
        \footnotesize
        \item[*] Low~power
        \item[**] Requirement
    \end{tablenotes}
    \end{threeparttable}
  \end{center}
\end{table*}

Considering similar application scenarios of the fast timing system discussed in this article, key parameters of the SiPM+PIST system as well as fast timing detectors at High-Luminosity LHC experiments are summarised in the Table~\ref{Table:summary}. It is noteworthy that the SiPM readout system with PIST ASIC exhibits excellent timing performance.

\section{Summary and prospects}
\label{sec:conclusion}
A SiPM-readout front-end chip named PIST was developed using $55~\mathrm{nm}$ CMOS technology to achieve picosecond-level fast timing for future electron-positron collider experiments. Comprehensive studies have been performed to characterise the fast timing performance for a dedicated PIST-SiPM test stand. Experimental results show that the time resolution of the SiPM-ASIC combined setup is $45~\mathrm{ps}$ at the 1-MIP signal level and can achieve better than $10~\mathrm{ps}$ with a large enough SiPM signal, while the intrinsic time resolution of the PIST ASIC is around $4.76~\mathrm{ps}$. With the ToT readout of the PIST chip, its output signal can be dramatically extended and is quantified in a range from $900$ to $40000$ p.e.. This low-power PIST chip can be can be a promising candidate in applications of SiPM-based detectors for fast timing at future Higgs factories.

\bibliographystyle{elsarticle-num} 

\end{document}